\documentclass[twocolumn,showpacs,preprintnumbers,amsmath,amssymb]{revtex4}

\usepackage{graphicx}
\usepackage{dcolumn}
\usepackage{bm}

\begin{document}

\preprint{}

\title{Cold SO$_2$ molecules by Stark deceleration}%

\author{O.~Bucicov}
\author{M.~Nowak}%
\author{S.~Jung}
\thanks{present address: NIST, Gaithersburg USA}
\author{E.~Tiemann}
\author{Ch.~Lisdat}
 \email{lisdat@iqo.uni-hannover.de}
\affiliation{Institut f\"ur Quantenoptik, Leibniz Universit\"at Hannover, Welfengarten 1, 30167 Hannover, Germany}%

\author{G.~Meijer}
\affiliation{Fritz-Haber-Institut der Max-Planck-Gesellschaft, Faradayweg 4--6, 14195 Berlin, Germany}%

\date{\today}

\begin{abstract}
We produce SO$_2$ molecules with a centre of mass velocity near zero using a Stark decelerator. Since the initial kinetic energy of the supersonic SO$_2$ molecular beam is high, and the removed kinetic energy per stage is small, 326 deceleration stages are necessary to bring SO$_2$ to a complete standstill, significantly more than in other experiments. We show that in such a decelerator possible loss due to coupling between the motional degrees of freedom must be considered. Experimental results are compared with 3D Monte-Carlo simulations and the quantum state selectivity of the Stark decelerator is demonstrated.
\end{abstract}

\maketitle

%
%
\section{Introduction}
\label{intro}
%
%
The production and control of cold molecules is a rapidly evolving field. Many techniques have been demonstrated to produce cold and ultracold molecules in a highly controlled way. However, it remains an open problem to produce large quantities of complex molecules of chemical interest. For these molecules Stark deceleration has proved to be a very efficient method to generate cold samples \cite{bet03,mee06a}. The decelerated molecules can be confined for long times \cite{hei07,saw07} and can be used for collision studies with precise control over the translational degrees of freedom \cite{gil06}.

 In this paper, we report the Stark deceleration of SO$_2$ from a supersonic beam. We have recently demonstrated that this technique can be successfully applied to this "unfavourable" molecule \cite{jun06}, which has a comparatively small ratio of Stark shift to initial kinetic energy and therefore requires a large number of deceleration stages. The application of Stark deceleration on similar molecules was also demonstrated before with YbF \cite{tar04}, with a different type of Stark decelerator \cite{bet02a}. 

The effort on SO$_2$ is motivated by the possibilities the molecule offers: it is known that the molecule can be dissociated at the threshold \cite{bec93} and that the dissociation process can be controlled by external fields \cite{jun06,jun06a}. Such field controlled reactions of cold molecules are frequently studied theoretically and are of significant interest \cite{abr07,hud06a,kre06,kre05,bal03a,avd03}. Additionally, the cold SO$_2$ fragments O and SO can serve as targets for further reaction studies. 

To explore these experiments high particle densities are required. To achieve this, one could accumulate the fragments (which have triplet ground states) in a magnetic trap by photodissociating the magnetic field insensitive, SO$_2$ within a magnetic trap. Decelerated SO$_2$ molecules can also serve as interesting collision partners, since the rotational structure is dense and thresholds for inelastic collisions will therefore appear at lower energy than in the previous study OH \cite{gil06}.

The article is organized as follows: in Sec.~\ref{dec} we summarize the operational principle of Stark deceleration and detail the theoretical background for modelling of the deceleration (Sec.~\ref{mod}). The experimental setup for SO$_2$ deceleration is outlined in Sec.~\ref{setup} and results of the measurements are presented in Sec.~\ref{meas}. We conclude in Sec.~\ref{conc} with a discussion of the possibilities for future trapping of SO$_2$.
%
%
\section{Stark deceleration}
\label{dec}
%
%
%
%
\subsection{Principle}
\label{princ}
%
%
\begin{figure}[t]
\begin{center}
\footnotesize
\resizebox{0.45\textwidth}{!}{  \includegraphics{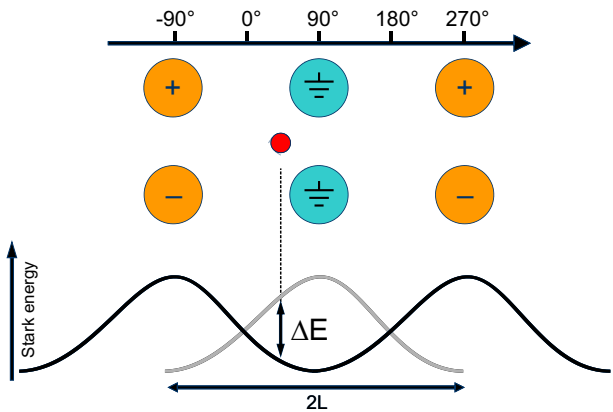}}
\caption{Schematic of the field configuration and potential energies shortly after switching the electric fields. The electrodes are symbolized by large circles, the polarity is indicated. The smaller circle depicts the synchronous particle.  The gray potential curve in the lower part of the figure gives the potential Stark energy before the field configuration is switched, the black curve thereafter. The potential energy loss of the molecule is depicted as $\Delta E$. The period of the decelerator is twice the stage separation $L$. At each switching, the potential is shifted by 1/2 of the period. The scale on top shows the position of a particle in the periodic structure of the decelerator in terms of the phase angle $\phi$. It can be seen that the synchronous particle is located at $\phi \approx 50^\circ$ at the switching time.}
\label{fig:schema}
\end{center}     
\end{figure}

The technique of Stark deceleration has been described in detail before \cite{bet99,mee05a,mee06} and only important points for this paper will be recalled here.

In Stark deceleration, the force generated by the interaction of the permanent electric dipole moment of the molecule and an inhomogeneous electric field is used to decelerate molecules. The inhomogeneous electric field is produced by a series of electrode pairs with opposing polarity, between which the molecular beam propagates. For molecules in an appropriate state (low field seeking) a potential well is generated. As the molecule enters a high-field
region, kinetic energy is removed. Leaving the field region the molecule would regain the kinetic energy, however, this is avoided by quickly grounding the electrodes, such that the potential energy $\Delta E$ is removed (see Fig.~\ref{fig:schema}). To deplete the total kinetic energy of the molecule several deceleration stages (each removing $\Delta E$) are required, for which the switching is repeated. With a modified geometry of electrodes and switching sequence, molecules in high field seeking states can also be decelerated \cite{bet02a,tar04}.

The energy $\Delta E$ removed from a molecule depends on the position of the molecule relative to the electrodes at the switching time. This position is described by the phase angle $\phi$, which has a period of twice the stage separation $L$. For the so called synchronous particle, the angle $\phi$ is constant throughout the deceleration process. For equally spaced electrode pairs, the time between subsequent switching events must, therefore, increase since the velocity of the molecule is lowered.  At $\phi = 90^\circ$ the synchronous molecule is right in-between a pair of electrodes when the field is switched. It loses the maximum amount of kinetic energy. For $\phi = 0^\circ$ it is in the middle between two pairs of electrodes and loses no energy (Fig.~\ref{fig:schema}).

Not only the synchronous molecule is decelerated, but also molecules that are close in phase space to the synchronous one. They form a so-called bunch. Molecules in a bunch are decelerated together due to the effect of phase stability. Phase stability can be visualised by considering e.g. a molecule, which is slightly ahead of the synchronous one and has the same velocity. This molecule will lose more kinetic energy at the next stage than the synchronous one since it traveled further up the potential hill, and will eventually fall behind the synchronous molecule. The inverse process will happen as for molecules behind the synchronous one. 

In addition to effects of phase stability, the molecules are guided through the decelerator by a focusing force toward the axis of the decelerator. This force stems from the increasing electric field strength toward the electrode surface. It was observed that an interplay between the bunching and guiding exists, which leads to unstable motion of the molecules in regions of phase space that would be regarded as stable from the 1D model described above \cite{mee05a,mee06}. For the modelling of the molecular trajectories in the decelerator, it is therefore necessary to perform full 3D calculations that reproduce these instabilities.  
%
%
\subsection{Modelling}
\label{mod}
%
%
\begin{figure}[t]
\begin{center}
\footnotesize
\resizebox{0.50\textwidth}{!}{  \includegraphics{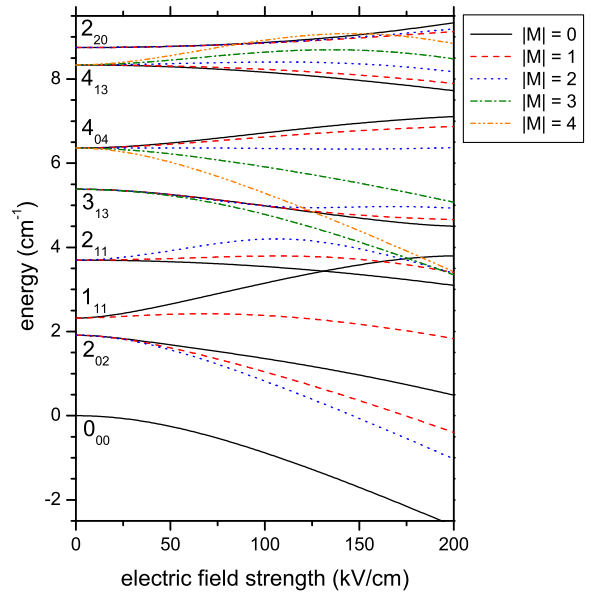}}
\caption{Stark shift of the lowest rotational lines $J_{K_- K_+}$ of the $v=0$ electronic ground state of SO$_2$. The maximum field strength on the molecular beam axis was about 90~kV/cm, thus well below the crossing of the levels $1_{11}$ and $2_{11}$, $M=0$ at about 130~kV/cm.}
\label{fig:stark}
\end{center}     
\end{figure}
To simulate the observed time-of-flight spectra, molecular trajectories through the decelerator are calculated using Monte-Carlo simulations where randomized starting velocities and times are generated for molecules in a chosen quantum state. These parameters are adapted to the experimental conditions (see Sec.~\ref{setup}). The classical equations of motion can then be integrated if the forces on the molecule are known. 

As mentioned above, the forces are due to the Stark interaction of the permanent electric dipole moment of the molecule (for SO$_2$ in the ground state about 1.6~D \cite{pat79}) with the inhomogeneous electric field produced by the electrodes. These forces can be calculated for a given field strength from the rotational constants of the molecule~\cite{mue00} and the dipole moment. The matrix elements necessary for the description of the rotational structure of an asymmetric top molecule and the calculation of the Stark shift of these levels are given e.g. in Ref.~\cite{jun06a}. The Stark shifts of the lowest rotational levels of the $v=0$ vibrational state of the $\widetilde{\rm X}$ $^1$A$_1$ state of SO$_2$ are depicted in Fig.~\ref{fig:stark}. The field configuration of the electrodes is calculated with a commercial program (simion, \cite{sim60}), the field of the hexapole lens in front of the decelerator can be derived from an analytical expression \cite{and97}.

%
%
\section{Setup}
\label{setup}
%
%
The decelerator apparatus we used is an extended and improved version of our Stark decelerator described before \cite{jun06}. The apparatus has had an additional 186 stages added and the mechanical stability has been improved. This apparatus is now able to stop SO$_2$ molecules. The pulsed SO$_2$ molecular beam is produced by expansion of 5\% of SO$_2$ in Xe as carrier gas through a 200~$\mu$m diameter nozzle into vacuum. By cooling the nozzle to 210~K the beam velocity is reduced to about 300~m/s. The pulse typically lasts 300~$\mu$s and has a rectangular profile as was confirmed by time-of-flight spectra. Due to the supersonic expansion the rotational temperature of the SO$_2$ molecules is reduced to about 6.5~K and most molecules are in the vibrational ground state.The velocity distribution is also compressed, and a longitudinal temperature of about 3.5~K is achieved.

The molecular beam passes after 2~cm through a skimmer with 1~mm diameter and reaches an ultra-high vacuum chamber. 5 cm behind the nozzle, an 8~cm long hexapole lens focuses molecules in the low field seeking states on the entrance of the Stark decelerator. The operation voltage of the hexapole is $\pm$11~kV and is optimized for the state $J_{K_- K_+} = 1_{11}$ $M = 0$ as is the Stark decelerator (see Sec.~\ref{quantum}).

The decelerator entrance is located 1.7~cm behind the hexapole. This spacing was optimized by trajectory simulations to match the molecular beam characteristics to the acceptance of the decelerator. The general setup of the Stark decelerator itself and the dimensions of the individual stages are comparable to other apparatuses \cite{bet03,bet00a}. The stages consist of two parallel, highly polished stainless steel electrodes with a diameter of 3~mm and a free gap of 2~mm. The centers of adjacent stages are separated by $L=5.5$~mm. In total, the decelerator consists of 326 stages, which are divided in two sets of 163 electrode pairs. Electrodes with the same polarity in one set are mounted in one bar, giving an electrode comb of about 1.8~m length. Deceleration stages from different sets interact with the molecular beam alternately. One set of electrode combs is oriented under a 90$^\circ$ angle with respect to the other set to guide the molecules in both transverse directions. Each electrode bar is mounted by three insulators, which are adjustable to align the spacing between the opposing electrode combs over the full length of the decelerator.

Four fast switches supply the high voltage for the electrode bars. The decelerator is currently operated at $\pm$10~kV, it is envisaged to rise the potential difference to 25~kV. The maximum field strength on the molecular beam axis was thus about 90~kV/cm. Up to this field strength, the molecular state $J_{K_- K_+} = 1_{11}$, $M=0$ experiences no avoided crossing with other levels, which could lead to particle losses due to adiabatic traversal of the crossing. This effect would dramatically change the guiding and bunching properties of the decelerator due to the changed Stark effect and will lead to particle loss with high probability. Such a crossing appears at about 130~kV/cm (see Fig.~\ref{fig:stark}), which will still not be experienced at an operation of the decelerator at 25~kV potential difference. Though the figure seems to indicate a real crossing of the two Stark curves and the states $1_{11}$ and $2_{11}$ are not directly coupled by the Stark Hamiltonian, the crossing is weakly avoided by indirect coupling via other rotational levels. Even if the crossing region would be reached, the particle losses are expected to be very small due to the small energy separation of a few $h*100$~Hz between the Stark curves: a Landau-Zener type model \cite{sch05i,wit05a} suggests for a molecular velocity of 60~m/s a probability of following the Stark curve adiabatically of less than 10$^{-6}$.

The loading current of the decelerator during the switching is limited by 1.5~k$\Omega$ high power resistors and is buffered by a 1~$\mu$F capacitor for each electrode comb. During one series of switching events (burst) to decelerate a bunch of molecules the voltage drops by about 3\%. The capacitors are reloaded by the power supply before the next burst, which follows at a 10~Hz rate.

The molecules can be detected 1~cm behind the nozzle and 4~cm behind the decelerator by laser induced fluorescence. In the first interaction zone, the density distribution of the molecules in the pulse is observed since the time-of-flight is too short to lead to a significant dispersion of different velocity classes. In contrast, the velocity dispersion is dominant in the second interaction zone. Hence the velocity distribution of the molecules in the pulse can be inferred from time-of-flight spectra there. The SO$_2$ molecules are excited by frequency doubled radiation from a pulsed coumarin-120 dye laser, which is pumped by a Xe:F excimer laser. For detection purposes we used the vibrational band (1,4,2) in the $\widetilde{\rm C}$ $^1$B$_2$ state \cite{jun06a,yam95,eba88}.
%
%
\section{Measurements}
\label{meas}
%
%
\begin{figure}[t]
\begin{center}
\footnotesize
\resizebox{0.50\textwidth}{!}{  \includegraphics{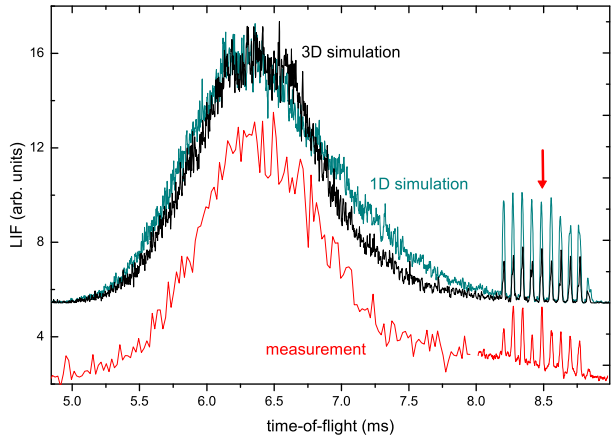}}
\caption{Measurement (lower curve) and simulations (upper curves) of time-of-flight spectra of SO$_2$ molecules in the rotational state $J_{K_- K_+} = 1_{11}$ $M = 0$ by laser induced fluorescence (LIF). The traces of measurements and simulations are shifted relative to each other for the sake of clarity. The two simulations compare 3D (darker curve) and 1D (lighter curve) trajectory calculations. The molecules are decelerated from 300~m/s to 157~m/s at a phase angle of $\phi=55^\circ$. The arrow marks the bunch with the synchronous molecule. The different signal-to-noise ratio in the two parts of the experimental data is due to longer averaging of the part with the decelerated bunches. (Clearly, a 1D simulation is not capable of correctly reproducing the measured intensities.)}
\label{fig:dec55}
\end{center}     
\end{figure}
%
%
%
\subsection{Deceleration and 3D effects}
\label{resdec}
%
%
As described in Sec.~\ref{setup} the decelerator setup is constructed for molecules in a low field seeking state and optimized for $J_{K_- K_+} = 1_{11}$ $M = 0$. To characterize the decelerator setup several measurements were performed and their results were compared with simulations. In Fig.~\ref{fig:dec55}, a time-of-flight spectrum of molecules in the vibrational ground state and $J_{K_- K_+}=1_{11}$ at the interaction zone behind the decelerator and corresponding simulations are depicted. A phase angle of $\phi = 55^\circ$ and an initial velocity of the synchronous particle of 300~m/s were chosen. This leads to a final velocity of 157~m/s for the synchronous molecule in the low-field-seeking component $M=0$ of the level $J_{K_- K_+}=1_{11}$. The initial mean velocity of the molecular pulse was determined to be 308~m/s from time-of-flight measurements in the guiding regime. For this measurement, a time independent voltage of $\pm$3.5~kV is applied to all electrodes to guide the molecules through the apparatus without deceleration. The signal is thus largely improved. The continuous operation is also preferable for the velocity measurement since no bunching of molecules appears.

Around a time-of-flight of 6.4~ms in Fig.~\ref{fig:dec55}, the non-decelerated part of the molecular pulse is visible. It consists of molecules in the $M=0$ component of the level $J_{K_- K_+}=1_{11}$, which do not occupy a region in phase space that forms a bunch of decelerated molecules. These molecules are partly guided through the decelerator and form the pulse in Fig.~\ref{fig:dec55}, while molecules in $\left| M \right| = 1$ are lost from the beam (see Sec.~\ref{quantum}). The nine peaks at an arrival time of about 8.5~ms are decelerated molecules. Several bunches are visible since the initial molecular pulse fills several adjacent stages of the decelerator and a train of bunches with different velocity is formed. The bunch with the synchronous molecule is marked by an arrow in Fig.~\ref{fig:dec55}.

The bunches experience a different sequence of time dependent electric fields and can thus finally have different velocities. Molecules that are considerably ahead of the synchronous molecule and are already in the decelerator when the switching of the high voltage starts, are decelerated as the marked synchronous particle, for which the switching starts when entering the decelerator. But these early molecules miss the last switching events, since they have left the decelerator before the synchronous particle and the end of the switching sequence. Thus they have finally a higher velocity than the synchronous molecule. This effect was also observed and investigated for molecules loaded into a storage ring \cite{cro01}. 

Molecules entering the decelerator later than the synchronous molecule miss the first switching events. The synchronous molecule is already slightly slowed down, when these molecules experience the electric field from the first stage. To match the switching sequence and the initial velocity of these later molecules, molecules with a lower initial velocity than the synchronous molecule form the bunches. These molecules leave the decelerator with the same velocity as the synchronous particle, but experience a longer field free flight to the detection zone since the switching of the electric field has stopped before the molecules leave the decelerator.
 
Figure~\ref{fig:dec55} shows the very good agreement between measured and calculated time-of-flight spectra (upper curves, 3D and 1D simulation). The agreement of the measured and simulated arrival times of significant structures in the time-of-flight spectrum indicates a good modelling of the general dimensions of the apparatus. More important is the good match of the relative intensities of the observed features with the 3D simulations, which show that the setup does not suffer from significant distortions of the field distribution in the decelerator. This is especially remarkable over the length of nearly two meters.

It is important to note that 3D simulations are essential to achieve this match \cite{mee06}. Omitting the coupling between longitudinal and transverse motion by 1D calculations produces molecular bunches that are about a factor of two to three more intense (see Fig.~\ref{fig:dec55}, 1D simulation). Closer investigations of the shape of the decelerated bunches showed characteristic structures as described by Meerakker {\it et al.} \cite{mee06}. It was difficult to find in our experiment an exact agreement between the observations and 3D simulations. The modelling of the deceleration was improved by including the 3\% drop of the electric potential during the burst, but no good agreement was found. We attribute this to remaining uncertainties in the geometry, which might be e.g. a slow variation of the electrode separation over the length of the decelerator. Such imperfections can in principle be modeled but raise the number of unknown parameters in the simulation unreasonably. Nevertheless, we can conclude that the deviations of the realized decelerator from the design geometry are small, since the general agreement between simulations and measurements is quite good.

\begin{figure}[t]
\begin{center}
\footnotesize
\resizebox{0.5\textwidth}{!}{  \includegraphics{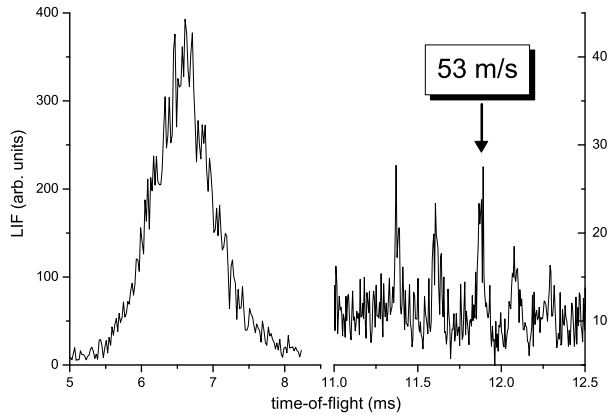}}
\caption{Time-of-flight spectra of SO$_2$ molecules in the rotational state $J_{K_- K_+} = 1_{11}$, $M = 0$. The molecules are decelerated from 285~m/s to 53~m/s (synchronous molecule, arrow) at an phase angle of $\phi=67^\circ$ and $\pm$10~kV.}
\label{fig:dec67}
\end{center}     
\end{figure}

Changing the phase angle influences the final velocity of the decelerated molecules and therefore their arrival time. Systematic investigations at different phase angles showed that the behaviour of the decelerator is well understood. In Fig.~\ref{fig:dec67} the recorded time-of-flight spectrum at a phase angle of $\phi = 67^{\circ}$ is shown. Here, molecules were decelerated from 285~m/s to 53~m/s. This final velocity is very close to typical velocities that are used to load electrostatic traps \cite{bet00,mee05,hei07}. A slight increase of the phase angle would have been sufficient to achieve this but the detection of the particles would have been very difficult due to dispersion of the bunches during the time-of-flight from the end of the decelerator to the interaction zone. The most probable velocity of the molecular pulse was about 300~m/s; the reduced initial particle number at 285~m/s was traded off against the lower kinetic energy.

Simulations of this experiment again predict the arrival times of the molecular pulses and the relative intensity of the decelerated bunches very well. Also, the increased dispersion of the bunches arriving late due to longer time-of-flight without electric fields is well described. Unfortunately, the simulations predict about 2.5 more decelerated molecules in comparison to the non-decelerated pulse. In our opinion, this can be either due to deviations in the description of the velocity distribution of the initial molecular pulse, collisional losses, or more likely due to imperfections in the modelling of the detection geometry, to which the results become very sensitive at low velocity.
%
%
\subsection{Quantum state selectivity}
\label{quantum}
%
%
An important aspect for the application of the decelerator setup in collision experiments is the quantum state selectivity of the apparatus. Beside the velocity selectivity, which depends on the phase angle chosen for deceleration (in the order of a few m/s), the quantum state selectivity is very important for the ability to observe and interpret state changing or reactive collisions.

Our decelerator for a molecule with comparatively dense rotational structure can show significantly different behaviour than shorter decelerators for lighter molecules with less dense rotational structure, since several low field seeking states will be initially populated (see Fig.~\ref{fig:stark}) and can be guided or even be decelerated. On the other hand, one can expect an improved removal of molecules in high field seeking or electric field insensitive states ($J_{K_- K_+} = 2_{11}$, $M=1$) due the longer interaction time with the high number of stages or simply by the divergence of the particle beam over the long flight path.

\begin{figure}[t]
\begin{center}
\footnotesize
\resizebox{0.50\textwidth}{!}{  \includegraphics{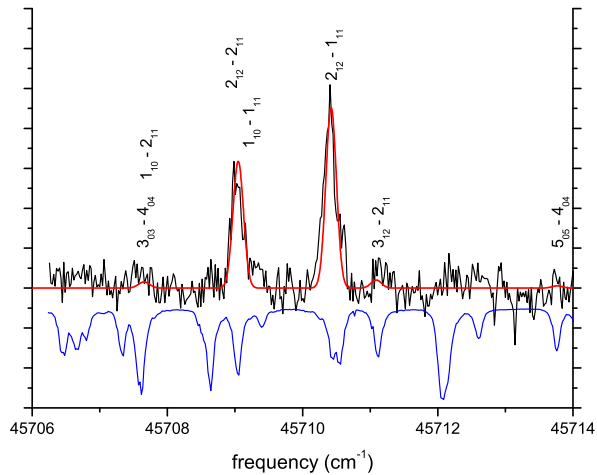}}
\caption{Measurement of the quantum state distribution behind the decelerator (black curve online) and Monte-Carlo simulation (red) for the main decelerated bunch (10~kV, $\phi = 55^{\circ}$). The inverted curve (blue) shows a reference spectrum before the decelerator with a thermal rotational distribution of about 6.5~K. Quantum numbers of observed and expected transitions are given.}
\label{fig:sel}
\end{center}     
\end{figure}

We have compared the spectra of the (1,4,2)-(0,0,0) band of SO$_2$ before and after the decelerator to investigate these aspects of quantum state selectivity. The decelerator was operated at $\pm$10~kV and a phase angle of 55$^{\circ}$, thus conditions leading to a time-of-flight spectrum are as depicted in Fig.~\ref{fig:dec55}. The laser for the spectroscopy of SO$_2$ was pulsed such that the molecules in the bunch indicated by the arrow in Fig.~\ref{fig:dec55} were excited. Tuning the laser frequency reveals a spectrum as shown in Fig.~\ref{fig:sel}.

The great simplification of the spectrum behind the decelerator with respect to a reference spectrum taken before the decelerator under the same experimental conditions (inverted curve in Fig.~\ref{fig:sel}) is obvious. Mainly two lines remain visible within the achieved signal-to-noise ratio. Both are transitions involving the population of the decelerated $J_{K_- K_+}=1_{11}$ $M = 0$ level. Since the region from the end of the decelerator to the detection zone is essentially field free, the orientation of the molecules in $M=0$ with respect to the local axis of quantization is lost. The population in the interaction zone is hence expected to be distributed homogeneously over all $M$-levels when probed with respect to a new axis of quantization introduced by the linear polarization of the laser. This loss of the quantization axis explains the intensity ratio of about 2:3 between both lines, since the transition with $M = 0$ is forbidden in $1_{10} - 1_{11}$ and saturation of the transitions is achieved, thus the H\"onl-London factor for the relative strength of both transitions has not to be taken into account. Hence, the line to $1_{10}$ only probes the population in $M=\pm1$, while the one to $2_{12}$ probes all three Stark levels of the ground state level $J_{K_- K_+}=1_{11}$.

The red curve in Fig.~\ref{fig:sel} is a simulation of the spectrum behind the decelerator. The relative intensities for the lines were taken from Monte-Carlo simulations of the decelerator for different molecular states but with the same switching sequence: the one optimized for the level $J_{K_- K_+}=1_{11}$ $M=0$. As can be seen in the figure, other transitions are expected of which also indications are visible in the measurements. These are due to other ground state levels with positive Stark effect like the level $2_{11}$ $M=2$ (see Fig.~\ref{fig:stark}). The simulations show that molecules in these levels are not decelerated since the smaller Stark effect and accordingly smaller energy loss $\Delta E$ per stage do not allow the formation of phase stable bunches with decelerated molecules. The weak transitions in Fig.\ref{fig:sel} show molecules in the low velocity tail of the initial molecular pulse. Only lines from low field seeking molecular levels are visible, since these molecules are guided independently from the exact switching sequence through the decelerator like the main pulses visible in Figs.~\ref{fig:dec55} and \ref{fig:dec67}. Molecules in high field seeking states are efficiently removed from the molecular beam.

The simulations tell us that the population in $J_{K_- K_+}= 1_{11}$ $M = 0$ is a factor of about 44 larger than in $J_{K_- K_+}= 2_{11}$ $M = 2$ or about 80 for $J_{K_- K_+}= 4_{04}$ $M = 0$. A precise limit for the population in high-field seeking states is difficult to derive, since even for several million simulated trajectories no particle reached the end of the decelerator. The suppression of other low field seeking quantum states can be easily improved by an increased time delay between the decelerated molecules and the guided main pulse. This does mostly mean higher deceleration by larger phase angles or voltages but not necessarily lower final velocity, since the initial molecular velocity can be increased easily by the choice of the nozzle temperature or the carrier gas. Therefore, a large parameter space can be covered for experiments with pure molecular samples.
%
%
%
\section{Conclusion}
\label{conc}
%
%
We have constructed a Stark decelerator with 326 stages for molecules in low field seeking states. The decelerator was successfully tested with SO$_2$ molecules and can be applied to other molecules with a small ratio of Stark shift to kinetic energy. The apparatus is expected to be capable of loading a trap for SO$_2$ molecules, which will be implemented soon. Improvements on the number of decelerated molecules are expected from an increase of the electric fields to a level achieved in our previous, shorter setup \cite{jun06}. The phase angle at which the decelerator is operated to produce slow molecules will be decreased, which increases the number of molecules in a decelerated bunch.

We have tested the performance of our apparatus and its experimental results by comparison of time-of-flight spectra obtained under different operation conditions of the decelerator with trajectory simulations. Good understanding of the experimental conditions in terms of the temporal sequence and the signal intensity was achieved. For the latter it is essential to include the transverse motion of the molecules in the decelerator.

The state selectivity of the apparatus was investigated for the dense level structure of the SO$_2$ molecule and good agreement was found with corresponding simulations. High state selectivity was achieved, the particle number in the next mostly populated quantum state at the arrival time of the synchronous molecule is a factor of 44 below the number of decelerated molecules in the designed quantum state. Molecules in high field seeking states are efficiently removed from the molecular beam. Improved state selectivity can be achieved by optimized operation conditions of the decelerator, i.e. stronger deceleration of the molecules or an improved beam source with a narrower velocity distribution.

These results show that a Stark decelerator of more than 300 stages is possible and can be applied for further experiments like studies of highly velocity selected collision or precision spectroscopy of cold molecules.

This work was supported by the German Science Foundation (DFG) within the priority program SPP~1116 and by the Research and Training Network Cold Molecules. We thank Nicolas Saquet for his support during his visits.

\end{document}